\documentclass[11pt]{article}
\pdfoutput=1

\usepackage{jheppub}
\usepackage{hyperref}
\usepackage{wrapfig}
\usepackage{amsmath}
\usepackage{amsthm}
\newtheorem{openproblem}{Open question}
\DeclareMathOperator{\Tr}{Tr}

\preprint{BRX-TH-6733}

\abstract{
As a consequence of the Ryu-Takayanagi formula, static semiclassical states in AdS quantum gravity share a set of special properties expressible as linear inequalities on their entanglement entropies. How general are these properties? Do they still hold when we move away from the regime of validity of RT, i.e. static states in classical Einstein gravity? Do they perhaps hold even for some class of non-holographic states? We will motivate these questions and review what's known about the answers.

\vspace{1cm}

\noindent This short article is based on a talk given by the second author at Strings 2026 in Shanghai and submitted as a contribution to \emph{Proceedings of Strings 2026}.
}

\title{Holographic entropy inequalities \& typical states}
\author{Guglielmo Grimaldi and Matthew Headrick} \affiliation{Martin Fisher School of Physics, Brandeis University, Waltham MA 02453, USA}

\emailAdd{ggrimaldi@brandeis.edu}
\emailAdd{headrick@brandeis.edu}

\begin{document} 

\maketitle

\section{Introduction}

\begin{figure}
    \centering
    \includegraphics[width=0.2\linewidth]{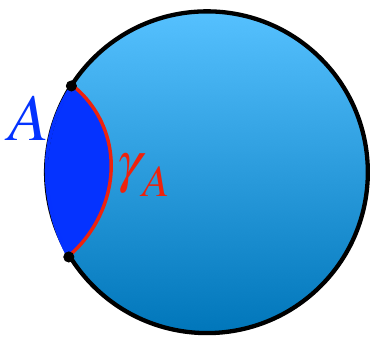}\hspace{1in}
    \includegraphics[width=0.21\linewidth]{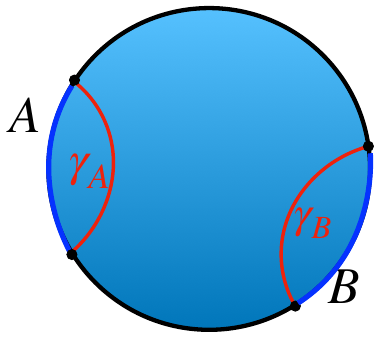}
    \caption{Left: Illustration of the homology condition in the RT formula \eqref{RT}. Right: In the holographic proof of subadditivity \eqref{SA}, the surface $\tilde\gamma_{AB}:=\gamma_A\cup \gamma_B$ is homologous to $AB$.}
    \label{fig:RTSA}
\end{figure}

Over the past 20 years, the Ryu-Takayanagi (RT) formula \cite{Ryu:06b20v, Ryu:2006ef} has taught us about important features of quantum gravity by revealing a direct connection between quantum information and spacetime geometry. Given a state $\rho$ of a quantum field theory and a spatial region $A$, we can define the von Neumann entropy $S_A := -\Tr(\rho_A\log\rho_A)$ of its reduced density matrix $\rho_A := \Tr_{A^{c}}(\rho)$. For holographic states, the RT formula states that the entanglement entropy $S_A$ of a spatial region $A$ on a boundary time slice is given by a geometric minimization problem:
\begin{equation}\label{RT}
S_A = \min_{\gamma\sim A}\frac{|\gamma|}{4G_{\rm N}}\,,
\end{equation}
where the minimum is taken over hypersurfaces $\gamma$ on the bulk slice that are homologous to $A$ (see figure \ref{fig:RTSA}, left).

An important point concerns the regime of validity of the prescription. The formula is valid when the holographic theory is at large $N$ and strong coupling, so that the bulk is well described by classical Einstein gravity, and when the state under consideration is static, or more precisely instantaneously static, which means it is 
an eigenstate of the time-reversal operator. (Not to be confused with a stationary state, which is an eigenstate of the Hamiltonian.) This implies that both the state and the region under consideration are invariant under time reversal. The existence of a moment of time symmetry means that we have a distinguished boundary Cauchy slice in the field theory and a corresponding distinguished bulk Cauchy slice in the gravitational theory.

All of the above assumptions can be relaxed, and prescriptions that allow for quantum corrections, higher-curvature corrections, and/or time-dependent states exist \cite{Faulkner:2013ana,Engelhardt:2014gca,Hubeny:2007xt,Dong:2013qoa}. In this article, we will first stay within the classical Einstein gravity regime, and then explore these extensions.

Entanglement entropies in quantum field theory are famously hard to compute in general. By making them easy to compute (subject to the above conditions), the RT formula has allowed us to learn about its features by acting as a keyhole into its behavior and properties, which would otherwise have been inaccessible. In this article, we argue that the RT formula still holds a whole set of mysteries waiting to be uncovered, which could teach us new lessons not only about holography but also about quantum field theory.

\section{Holographic entropy inequalities}

We are interested in universal properties of holographic entanglement entropy, meaning those that hold regardless of the details of the bulk spacetime and boundary region, as long as they are within the regime of validity of the RT formula. Of particular interest are linear inequalities involving the entropies of multiple boundary regions, known as holographic entropy inequalities (HEIs). The simplest example is \emph{subadditivity} (SA):
\begin{equation}\label{SA}
    S_{A} + S_B \geq S_{AB}\,.
\end{equation}
This inequality can be expressed as the non-negativity of mutual information, $I_{A:B} := S_{A} + S_B - S_{AB} \geq 0$. Its holographic proof is straightforward. Consider the two RT surfaces $\gamma_A$ and $\gamma_B$, whose areas compute $S_A$ and $S_B$, respectively. The surface defined by their union, $\tilde{\gamma}_{AB} := \gamma_A \cup \gamma_B$, is a \emph{candidate} surface for the computation of $S_{AB}$, since it is homologous to the region $AB$, via the union of the respective homology regions for $A$, $B$ (see figure \ref{fig:RTSA}, right). However, it is not necessarily the minimal surface, and therefore need not coincide with the true RT surface $\gamma_{AB}$. This gives us the desired bound:
\begin{equation}
    S_A + S_B = \frac{|\gamma_{A}|}{4G_{\sf{N}}} + \frac{|\gamma_{B}|}{4G_{\sf{N}}} = \frac{|\tilde{\gamma}_{AB}|}{4G_{\sf{N}}} \geq \frac{|\gamma_{AB}|}{4G_{\sf{N}}} = S_{AB}\,.
\end{equation}

\begin{figure}
    \centering
    \includegraphics[width=\linewidth]{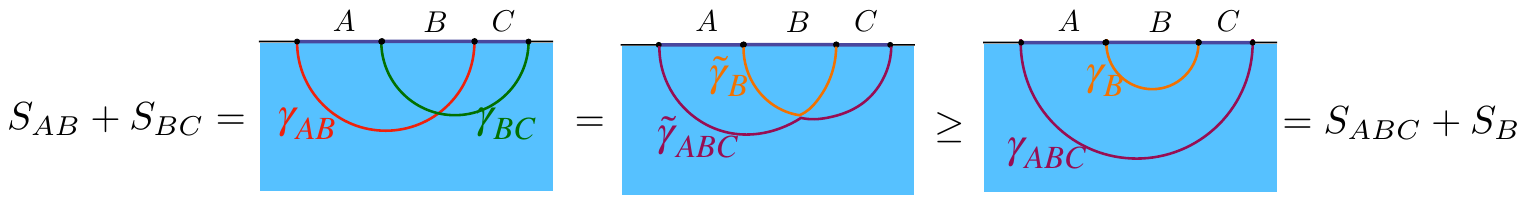}
    \caption{Holographic proof of strong subadditivity \eqref{SSA}.}
    \label{fig:ssaproof}
\end{figure}

Using the same method, we can prove more complicated inequalities, such as \emph{strong subadditivity} (SSA):
\begin{equation}\label{SSA}
    S_{AB} + S_{BC} \geq S_{B} + S_{ABC}\,.
\end{equation}
This inequality can also be written in terms of mutual information, as the monotonicity property $I_{A:BC} \geq I_{A:B}$. The proof is conceptually identical to that of SA, although slightly more involved \cite{Headrick:2007km,Headrick:2013zda}. We begin by considering the RT surfaces $\gamma_{AB}$ and $\gamma_{BC}$. We then construct two candidate surfaces, $\tilde{\gamma}_B$ and $\tilde{\gamma}_{ABC}$, by appropriately cutting and gluing the two surfaces $\gamma_{AB}$ and $\gamma_{BC}$, as shown in the figure. Once again, the areas of the two candidate surfaces provide upper bounds on those of the true minimal surfaces, thus proving the inequality. (See figure \ref{fig:ssaproof}.)

Note that, beyond the assumption of the validity of the RT formula, no additional physics enters the proofs of these inequalities. In particular, no information about the bulk equations of motion, energy conditions, or boundary conditions is required. In other words, the proofs go through if we work with any manifold-with-boundary equipped with a Riemannian metric.

Following this proof strategy, one can establish an infinite family of linear inequalities obeyed by the RT formula. One of the simplest is \emph{monogamy of mutual information} (MMI) \cite{Hayden:2011ag,Headrick:2013zda}:
\begin{equation}
    S_{AB} + S_{BC} + S_{AC} \geq S_A + S_B + S_C + S_{ABC}\,,
\end{equation}
which can also be expressed in terms of mutual information as the superadditivity property $I_{A:BC} \geq I_{A:B} + I_{A:C}$. Once again, the proof follows the same general structure. A possible information-theoretic explanation for the MMI inequality was conjectured in \cite{Cui:2018dyq}.

The systematic search for HEIs was initiated by the authors of \cite{Bao:2015bfa}. The clever idea of their work was to translate the problem into graph theory. Given a fixed geometry and a set of $\sf{N}$ boundary regions, one can draw all the RT surfaces associated with both fundamental \emph{and} composite boundary regions. Together, these surfaces partition the bulk slice into codimension-0 regions, which we associate with vertices of a graph. We then join vertices that share a portion of an RT surface by an edge whose weight is given by the area of that portion. We have thus obtained a weighted graph, and, by construction, the RT entropies are given by min-cuts on this graph. The converse is also true: starting from any weighted graph, one can construct a bulk geometry whose RT entropies reproduce the graph's min-cuts. Therefore, the problem of studying HEIs is essentially a problem in discrete mathematics and graph theory. We stress once again that their proofs involve almost no physical input: one may think of these inequalities as kinematic statements of holography.

The formal object that captures all of these inequalities is the so-called \emph{holographic entropy cone} (HEC), defined as follows. Given $\sf{N}$ external vertices, the min-cut function $\mathsf{S}$ is a function on $2^{[\sf{N}]}\setminus\{\emptyset\}$. We can collect its values into an \emph{entropy vector},
\begin{equation}
    \vec{\mathsf{S}} :=
    (S_1,\, S_2,\, S_3,\, \dots,\, S_{12},\, S_{13},\, \dots,\, S_{12\dots\sf{N}}).
\end{equation}
The set of all allowed entropy vectors forms the HEC for $\sf{N}$ parties. The HEC is a convex polyhedral cone, meaning it is defined by finitely many linear inequalities \cite{Bao:2015bfa}. Each facet of the cone is defined by the saturation of one of these inequalities.

For $\sf{N} \leq 4$ boundary regions, there are no additional inequalities beyond SA, SSA, and MMI. For $\sf{N} = 5$, there are 5 additional inequalities, for a total of 8 \cite{HernandezCuenca:2019wgh}. For $\sf{N} = 6$, more than 80,000 inequalities have been found \cite{Hernandez-Cuenca:2023iqh, hecdatacurrent}. Furthermore, two infinite families valid for arbitrary $\sf{N}$ are known \cite{Czech:2022fzb, Czech:2024rco}. Some progress has been made toward understanding the general structure governing these inequalities \cite{Hubeny:2018trv, Hubeny:2018ijt, He:2019ttu,He:2020xuo,Avis:2021xnz}.

\section{Beyond holography}

Except for SA, SSA, and MMI, the HEIs do not admit simple expressions in terms of mutual information, at least not ones that make their meaning immediately clear. Extracting a quantum-information-theoretic interpretation for the higher HEIs remains an open question. Some work on this problem includes \cite{Hernandez-Cuenca:2023iqh, Czech:2023xed,Czech:2026tgj,Czech:2025tds}. Considering the classes of states that obey them may help us better understand this question.

SA and SSA are valid entropy inequalities for general quantum states. All the remaining HEIs are \emph{not} valid for general quantum states. In this sense, higher HEIs provide necessary conditions for a quantum state to admit a classical dual geometry. Examples of states that violate them include the GHZ state for 4 or more parties, the ground state of a free massive scalar field, and, of course, states incorporating $1/N$ bulk quantum corrections, such as those arising from bulk quantum fields. These small quantum corrections will not push an entropy vector lying in the interior of the holographic entropy cone outside it, but they can push those lying on its boundary out. Therefore, one may think of $1/N$ corrections as making the holographic entropy cone ``fuzzy'' near its boundary.

Clearly, then, these inequalities do not provide universal statements about the entanglement entropies of quantum systems. However, it turns out that they hold for a much broader class of states than just classical, static holographic states. For example, any state with RT-like entropies will obey these inequalities. Examples include static asymptotically flat spacetimes connected by wormholes \cite{Gupta:2025jlq}, perfect-tensor states, random states of large systems, random tensor networks \cite{Hayden:2016cfa}, and measured graph states \cite{Jeffrey:2026oyb}. In all these examples, the entropies are computed through some form of minimization procedure similar to the RT formula, and therefore the RT proofs of the HEIs apply to them.

There are also other examples that do not obey an RT-like formula and yet appear to satisfy all of the HEIs. These possibilities are more speculative, owing to the scarcity of explicit examples, but are arguably even more interesting because of their weaker connection to the RT formula. They include perturbative $\alpha'$ corrections to the RT formula (corresponding to higher-derivative gravity theories) \cite{Dong:2013qoa}, ground states of large-$c$ CFTs with vacuum-block dominance (e.g.\ symmetric-product orbifolds) \cite{Hartman:2013mia}, ground states of topological theories \cite{Levin_2006,Kitaev:2005dm}, and ground and thermal states of sparse SYK models (which have no spatial dimensions, so the entropies are those of groups of fermions computed in flavor space) \cite{Iizuka:2024die}.

\begin{figure}
    \centering
    \includegraphics[width=0.5\linewidth]{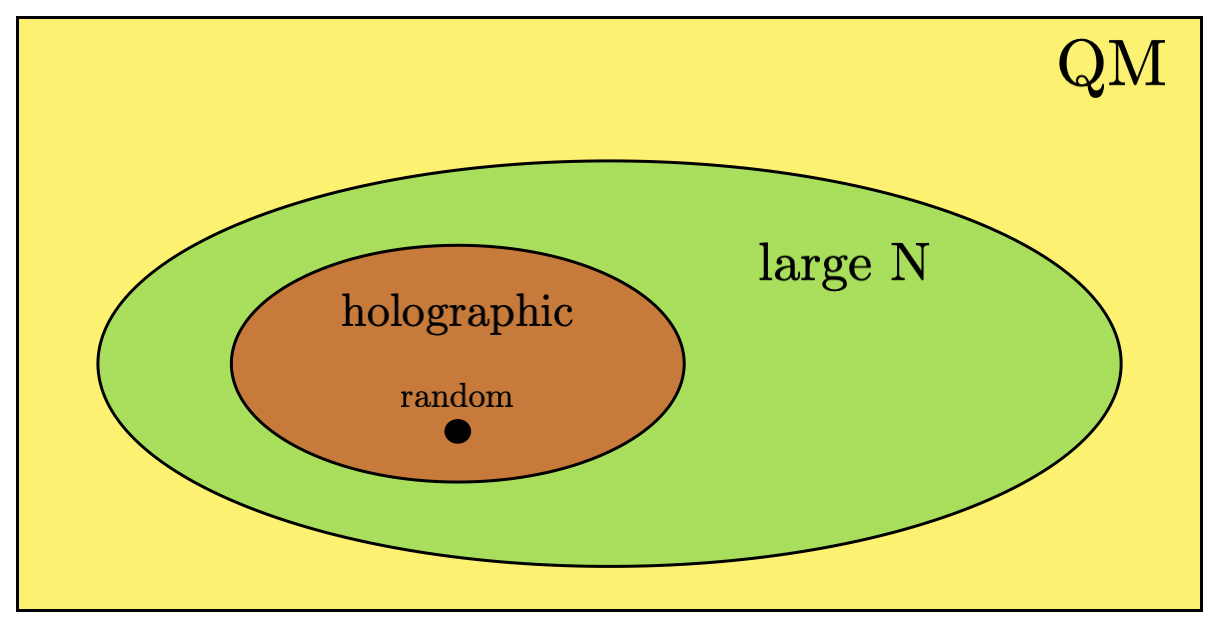}
    \caption{Some nested classes of states. We know that the HEIs apply to all holographic states, which includes random states, but not to all quantum states. Do they apply to some class of ``large-$N$'' states, suitably defined?}
    \label{fig:largeN}
\end{figure}

Many of the examples of HEI-obeying systems are ones with a large number of degrees of freedom. This raises the intriguing possibility that the HEIs are not fundamentally properties of holographic states but rather of ``typical'' (in some as-yet undefined sense) states of large-$N$ systems (see figure \ref{fig:largeN}).

\section{Beyond static states}

We can generalize the RT formula in a different direction by staying within the confines of classical gravity but allowing non-static (i.e.\ time-dependent) states. A sharp question about HEIs is whether they are obeyed for such states. For non-static classical holographic states, the RT formula is generalized to the Hubeny-Rangamani-Takayanagi (HRT) formula. Here, we are instructed to find the set of \emph{extremal}, codimension-2 achronal surfaces in the spacetime that are homologous to the region $A$, and select the one with the smallest area. The question then becomes: does the HRT formula obey the same set of linear constraints as the RT formula?

The proof of SA goes through unchanged for HRT, since the surface $\tilde\gamma_{AB}:=\gamma_A\cup\gamma_B$ is extremal and homologous to $AB$, and is therefore within the class of surfaces we minimize over.
However, we already encounter difficulties with SSA. The surfaces $\gamma_{AB}$ and $\gamma_{BC}$ are codimension-2 surfaces, so they will not generically intersect in the bulk. Even if they did intersect along a fully codimension-3 locus so they could be cut and glued to make candidate surfaces $\tilde{\gamma}_{B},\tilde{\gamma}_{ABC}$ in the correct homology classes, those surfaces would not be extremal, and hence would not belong to the set of surfaces over which we minimize. So proving that HRT obeys SSA, an important consistency check on the formula, is a non-trivial problem. Nonetheless, it has indeed been proven using multiple different strategies \cite{Wall:2012uf,Bousso:2024ysg, Grado-White:2025jci}. What is even more interesting is that these proofs, unlike the very simple one for RT, are full-fledged general relativity theorems, requiring AdS boundary conditions, global hyperbolicity, field equations, and energy conditions. The quantum information encoded in general relativity through holography thus knows something about the physics of the bulk.

The validity of higher HEIs for the HRT formula remains an open problem. One possible proof strategy would be to construct a graph model: a weighted graph on which the min cuts compute HRT entropies \cite{Grado-White:2025jci}. In addition to proving that the HEIs apply to time-dependent holographic states, such a graph model would be of some interest in its own right. For example, it would clarify the construction and interpretation of tensor-network models of time-dependent holographic states.

While we do not have a general proof, there is already substantial evidence that the HEIs do hold for time-dependent states. This evidence, including both partial proofs and numerical searches for counterexamples, spans a wide variety of different regimes:
\begin{enumerate}
    \item A general proof of MMI \cite{Wall:2012uf, Bousso:2024ysg}.
    \item A proof that all HEIs hold in the scaling limit of late times and large distances, using the membrane theory of entanglement dynamics \cite{Bao:2018wwd}.
    \item A proof that all HEIs hold in $2+1$ bulk dimensions when the bulk is either simply connected or a two-boundary genus-zero wormhole \cite{Czech:2019lps, Grado-White:2024gtx}.
    \item Numerical checks of thousands of known HEIs on thousands of randomly generated $2+1$ dimensional vacuum multiboundary wormhole configurations \cite{Grado-White:2024gtx}.
    \item A proof that no HEI can be violated by deforming a large class of saturating configurations \cite{Grimaldi:2025jad, Grimaldi:2026lbq,Czech:2026zca}.
\end{enumerate}

\begin{figure}
    \centering
    \includegraphics[width=0.3\linewidth]{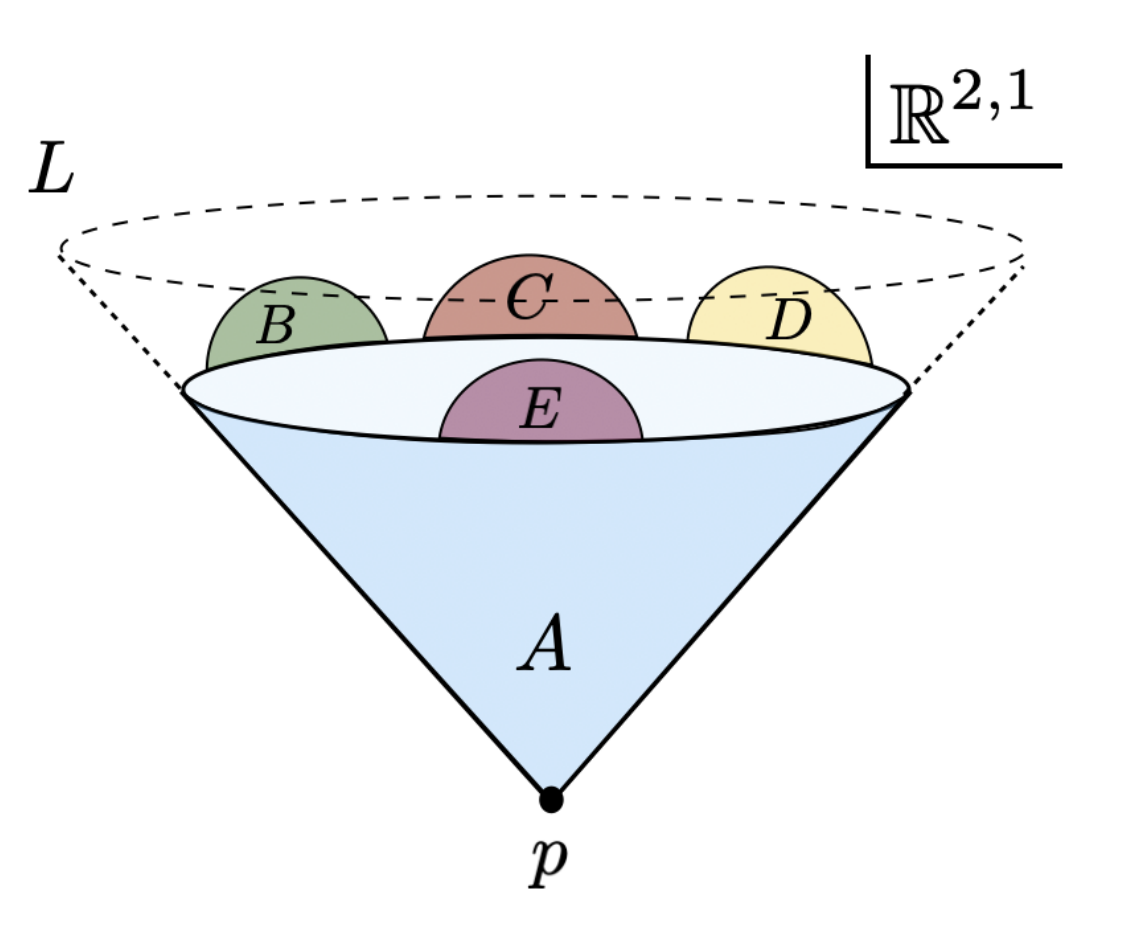}\hspace{1in}
     \includegraphics[width=0.25\linewidth]{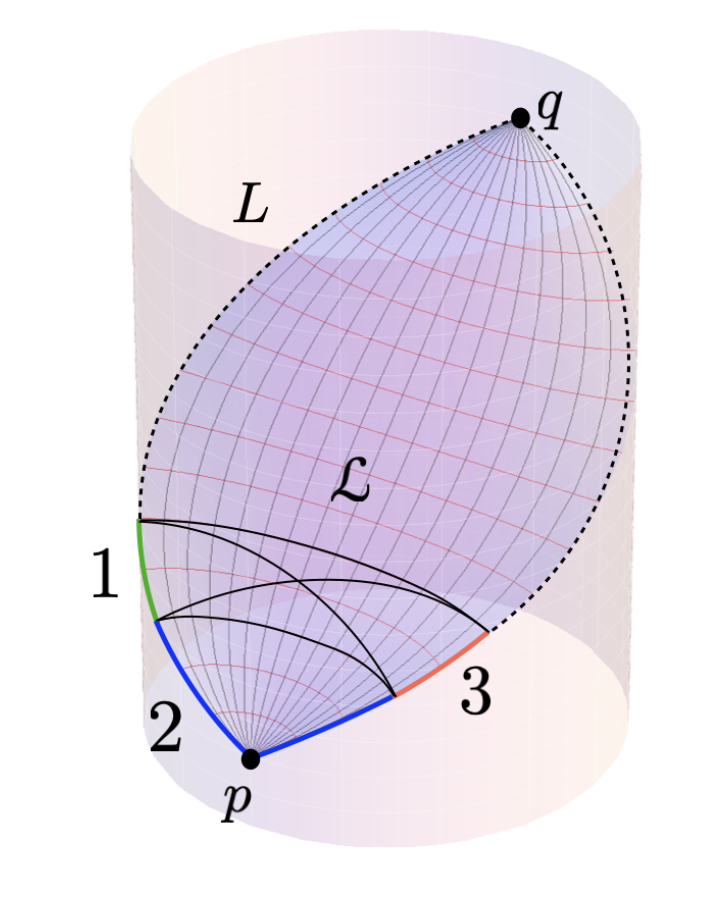}
    \caption{Left: Example of a light-cone configuration in $\mathbb{R}^{2,1}$. Right: Saturation of SSA on a light-cone configuration from the bulk perspective: every HRT surface lies on the Poincare horizon, and every null geodesic crosses every HRT surface once.}
    \label{fig:lcc}
\end{figure}

In the rest of the talk, we will discuss this last result. Given an HEI, we begin by arranging all the relevant parties in a light-cone configuration. This particular configuration is chosen because any entropy terms in the inequality that do not contain the central region cancel in any state, reducing the inequality to what is known as a \emph{null reduction}. For example, the inequality
\begin{equation}
    S_{ABC} + S_{BCD} + S_{CDE} + S_{ADE} + S_{ABE} \geq S_{AB} + S_{BC} + S_{CD} + S_{DE} + S_{AE} + S_{ABCDE}
\end{equation}
when evaluated on the light-cone configuration shown in figure \ref{fig:lcc} (left), reduces to
\begin{equation}
    S_{ABC} + S_{ADE} + S_{ABE} \geq S_{AB} + S_{AE} + S_{ABCDE}.
\end{equation}
If we consider the vacuum state, the reduced state on the remaining regions appearing in the null reduction is a so-called \emph{generalized Markov state} \cite{Casini:2017roe}. This implies that the inequality is saturated in the vacuum. This saturation can be easily understood from the bulk perspective, as shown in figure \ref{fig:lcc} (right). All the HRT surfaces lie on a common bulk light cone with vanishing expansion, so the area element remains constant along any geodesic. Thus, each geodesic contributes to the null-reduced inequality an amount proportional to the sum of all the coefficients in the inequality, which can be shown to vanish (as a consequence of a combinatorial property of HEIs known as \emph{balance}).

Having established that the vacuum state saturates the inequality, we can now perturb the state slightly. The perturbation must be compatible with the null energy condition, which implies focusing. Focusing, in turn, implies that along any null generator, the change in area must be a concave function of the affine parameter. It can be shown that focusing protects the validity of the inequality under such perturbations if the inequality itself satisfies a combinatorial property known as \emph{positive combinations majorization} (PCM). This property can be proven to hold for all holographic entropy inequalities \cite{Grimaldi:2026lbq,Czech:2026zca}. The connection between the dynamics of gravity and the combinatorial properties of these inequalities is remarkable. We therefore conclude that no HEI can be violated by deforming a light-cone configuration.

\section{Conclusion}

We have discussed four important questions surrounding HEIs:

\begin{openproblem}
What is the general structure or pattern of these inequalities?
\end{openproblem}

\begin{openproblem}
What do they tell us about the nature of entanglement in semiclassical states of quantum gravity? Is there an ansatz for such states that automatically satisfies them? 
\end{openproblem}

\begin{openproblem}
Are they valid beyond the Einstein gravity (strong-coupling) limit? What about for an even larger class of large-$N$ states?
\end{openproblem}

\begin{openproblem}
Are they valid for non-static states?
\end{openproblem}

We encourage all interested readers to pursue these problems. Solving them promises to teach us new and remarkable things about holography, classical and quantum gravity, and quantum systems in general.

\acknowledgments{The work of G.G. was supported by the U.S. Department of Energy through award DE-SC0009986. The work of M.H. was supported by the U.S. Department of Energy through award DE-SC0009986, and by a Simons Sabbatical Visiting Chair at the Institut des Hautes Etudes Scientifiques.}

\bibliographystyle{jhep}
\bibliography{references}
\end{document}